\documentclass[a4paper,dvips,12pt]{article}
\input{axodraw.sty}
\usepackage{amsmath,amssymb,exscale}   
\usepackage{array,multicol}
\usepackage{afterpage,float,flafter}   
\usepackage{epsfig,rotating,pifont,fancybox}   
\makeatletter   
\@addtoreset{equation}{section}   
\makeatother   
   
\newcommand{\bea}{\begin{eqnarray}}   
\newcommand{\eea}{\end{eqnarray}}   
   
\newcommand{\NPB}[3]{\emph{ Nucl.~Phys.} \textbf{B#1} (#2) #3}   
\newcommand{\PLB}[3]{\emph{ Phys.~Lett.} \textbf{B#1} (#2) #3}   
\newcommand{\PRD}[3]{\emph{ Phys.~Rev.} \textbf{D#1} (#2) #3}   
\newcommand{\PRL}[3]{\emph{ Phys.~Rev.~Lett.} \textbf{#1} (#2) #3}   
\newcommand{\ZPC}[3]{\emph{ Z.~Phys.} \textbf{C#1} (#2) #3}   
\newcommand{\PTP}[3]{\emph{ Prog.~Theor.~Phys.} \textbf{#1}  (#2) #3}

\newcommand{\JHEP}[3]{\emph{ JHEP} \textbf{#1} (#2) #3}

\def\simlt{\stackrel{<}{{}_\sim}}
\def\simgt{\stackrel{>}{{}_\sim}}
\title{   
\vspace*{-0.8cm}   
\begin{flushright}   
\normalsize{      
IEM-FT-212/01\\
IFT-UAM/CSIC-01-09\\   
\texttt{hep-ph/0103058}}\\ 
\end{flushright}    
\vspace{1cm}
\Large\textbf{Supersymmetry and Finite Radiative Electroweak Breaking from  
an Extra Dimension~\footnote{Work 
supported in part by CICYT, Spain, under contract AEN98-0816,
and by EU under contracts HPRN-CT-2000-00152 and HPRN-CT-2000-00148.}}
\vspace*{.5cm}
\author{\large
{\bf A.~Delgado and M.~Quir{\'o}s}\\ \\
\emph{Instituto de Estructura de la Materia (CSIC), Serrano 123,}\\
\emph{E-28006-Madrid, Spain.}}}
\date{}   
\begin{document}
\maketitle
\thispagestyle{empty}
\vspace*{.5cm}

\begin{abstract}\noindent
A five dimensional $N=1$ supersymmetric theory compactified on the orbifold 
$S^1/\mathbb{Z}_2$ is constructed. Gauge fields and $SU(2)_L$ singlets 
propagate in the bulk ($U$-states) while $SU(2) _L$ doublets are localized 
at an orbifold fixed point brane ($T$-states). Zero bulk modes and localized 
states constitute the MSSM and massive modes are arranged into $N=2$ 
supermultiplets. Superpotential interactions on the brane are of the 
type $UTT$. Supersymmetry is broken in the bulk by a Scherk-Schwarz mechanism 
using the $U(1)_R$ global $R$-symmetry. A radiative finite electroweak 
breaking is triggered by the top-quark/squark multiplet $\mathbb{T}$ 
propagating in the bulk. The compactification radius $R$ is fixed by the 
minimization conditions and constrained to be $1/R\simlt 10-15$ TeV. 
It is also constrained by precision electroweak measurements to be 
$1/R\simgt 4$ TeV. The pattern of supersymmetric mass spectrum is well 
defined. In particular, the lightest supersymmetric particle is the sneutrino 
and the next to lightest supersymmetric particle the charged slepton, 
with a squared-mass difference $\sim M_Z^2$. The theory 
couplings, gauge and Yukawa, remain perturbative up to scales $E$ given, 
at one-loop, by $ER\simlt 30-40$. Finally, LEP searches on the MSSM 
Higgs sector imply an absolute lower bound on the SM-like Higgs mass,
around 145 GeV in the one-loop approximation.
\end{abstract}
\vspace{2.cm}   
   
\begin{flushleft}   
March 2001 \\   
\end{flushleft}   
\newpage

%\begin{multicols}{2}
\section{Introduction} 
\label{introduction}
The Higgs boson is, for the time being, the only missing ingredient of the 
Standard Model (SM) of electroweak and strong interactions and, by far, the
most intriguing one. While it is related to the origin of gauge 
boson and fermion 
masses, the mechanism of electroweak breaking is intimately related to the 
so-called hierarchy problem which has given rise to the (minimal)
supersymmetric extension (MSSM) of the SM. In particular the 
radiative corrections to the squared Higgs 
mass in the SM have a quadratic sensitivity to the cutoff of the theory, 
$\Lambda_s$, which destabilizes the Higgs mass 
towards the region where the SM is no
longer reliable~\cite{quadratic}. 
This behaviour is softened in the MSSM where the sensitivity 
to the SM cutoff is only logarithmic and can therefore be interpreted as the
renormalization group running from the scale $\Lambda_s$ to the 
weak scale~\cite{logarithmic}. Actually, one
of the great successes of the MSSM is that the squared Higgs mass term can be
driven by radiative corrections generated by 
the top Yukawa coupling from positive values at the scale $\Lambda_s$ 
to negative values at the weak scale thus triggering radiative electroweak 
breaking~\cite{radiative}. 
Still the MSSM shows some (logarithmic) sensitivity
to the cutoff scale $\Lambda_s$,
whose value controls the total evolution of the squared Higgs mass. 

The sensitivity of the squared Higgs mass term on the cutoff $\Lambda_s$ 
through radiative corrections can
still be softened if the MSSM, and in particular the top/stop sector, is
living in the bulk of an extra dimensional space of size
$\mathcal{O}({\rm TeV}^{-1})$ \cite{large}. In that case the
radiative corrections to the squared Higgs mass term are not sensitive at 
all to $\Lambda_s$. In fact they are
finite, controlled by the inverse radius $1/R$ of the compactified extra 
dimensions~\footnote{This very well known fact in ordinary field
theory at finite temperature $T$, i.e. compactified on the circle of inverse
radius $T$, is at the origin of the so-called thermal (Debye) masses.}, and
with a sign which depends on the spin of the bulk particle circulating in the
loop~\cite{peskin,savas,alex1}. 
This observation gave rise to proposing the top/stop (hyper)multiplet
living in the bulk~\cite{alex1} as the source of a finite electroweak 
radiative breaking~\footnote{See footnote 11 in Ref.~\cite{alex1}.}, while some
explicit examples along that direction have been recently 
proposed in the frameworks of string~\cite{ignas} and field
theory in higher dimensions~\cite{Barbieri,Hall}.

Another issue which is highly related to the hierarchy problem and electroweak
breaking is that of supersymmetry breaking. The scale of
supersymmetry breaking must not be hierarchically different from the weak
scale since, on the one hand, we do not want to re-create the hierarchy 
problem, and on the other hand, it should trigger electroweak breaking. 
Because, in the finite radiative electroweak breaking, the weak scale is
provided (apart from loop factors) by the inverse radius $1/R$ of the 
compactified extra dimensions, that is the expected order of magnitude for
the scale of supersymmetry breaking. Although there are several mechanisms in
the literature which can provide the correct order of magnitude for
supersymmetry breaking, the one that naturally leads to supersymmetry 
breaking size of order $1/R$ is the Scherk-Schwarz (SS) 
mechanism~\cite{SS}-\cite{PQ}. In fact
both recent examples of finite radiative electroweak 
breaking~\cite{Barbieri,Hall} use, among other mechanisms, a variant of the
SS-mechanism based on a \emph{discrete} symmetry of the supersymmetric theory,
the $R$-parity. 

In this paper we will analyze a very simple 
five dimensional (5D) model where finite electroweak
breaking is triggered by the top/stop multiplet living in the bulk of the
extra dimension and supersymmetry is broken by a SS-mechanism based on a
\emph{continuous} symmetry of the 5D supersymmetric theory, $SU(2)_R$. So, 
unlike Refs.~\cite{Barbieri,Hall} supersymmetry breaking is controlled
by a continuous parameter, and the supersymmetric limit is 
continuously attainable. In this
sense, and although the theoretical setup of our 5D theory is rather
different from those presented in Refs.~\cite{Barbieri,Hall}, our results can
be considered in some aspects as more general than theirs.

The outline of this paper is as follows. In section 2 we will present the model
and the mechanism of supersymmetry breaking. Finite radiative electroweak 
breaking will be analyzed in section 3 and the Higgs sector and 
supersymmetric spectrum will be presented in sections 4 and 5, respectively. 
In section 6 a discussion on unification and non-perturbativity scales will
be done and some comments concerning the relation of our
paper with Refs.~\cite{Barbieri,Hall} will be made. 
Finally in section 7 we will 
present our conclusions and comparison with recent related works. 

\section{The 5D MSSM and supersymmetry breaking}
\label{model}

In this section we will describe a 5D $N=1$ model whose massless modes 
constitute the usual four dimensional
(4D) $N=1$ MSSM, where supersymmetry breaking is a bulk
phenomenon induced by the SS-mechanism, 
and with finite radiative electroweak breaking
triggered by the presence of a bulk top/stop hypermultiplet.

The 5D space-time is compactified on $\mathcal{M}_4\times S^1/\mathbb{Z}_2$,
where the $\mathbb{Z}_2$ parity is acting on the fifth coordinate as
$x_5\to -x_5$. The orbifold $S^1/\mathbb{Z}_2$ has two fixed points at
$x_5=0,\pi R$ and the compactified space has two 3-branes located at
the fixed points of the orbifold. In this way fields in the theory can be of
two types: those living in the 5D bulk, similar to untwisted states in
the heterotic string language ($U$-states), 
and those living on the branes localized at the
fixed points of the orbifold, similar to the heterotic string twisted states
($T$-states). 
We will assume for simplicity that $T$-states are localized at the
$x_5=0$ fixed point. While $U$-states feel the fifth dimension, i.e. their
wave function depend on $x_5$, $T$-states do not.

Vector fields live in the bulk and they are in $N=2$ vector multiplets 
in the adjoint
representation of the gauge group $SU(3)\times SU(2)_L\times U(1)_Y$,
$\mathbb{V}=(A_\mu,\lambda_1;\Phi,\lambda_2)$~\footnote{Where $i=1,2$
transform as $SU(2)_R$ indices and the complex scalar $\Phi$ is
defined as, $\Phi\equiv \Sigma+i\, A_5$.}.
Matter fields in the bulk are arranged in $N=2$
hypermultiplets, 
$\mathbb{H}=(\widetilde\psi_R,\psi_R;\widetilde\psi_L,\psi_L)$~\footnote{
In fact, $(\widetilde\psi_R,\widetilde\psi_L)$ transforms as a doublet 
under $SU(2)_R$.}. 
$\mathbb{Z}_2$, the parity in the fifth dimension, has
an appropriate lifting to spinor and $SU(2)_R$ indices, such that we can
decompose $\mathbb{V}$ and $\mathbb{H}$ into \emph{even},
$(A_\mu,\lambda_1)$ and $(\widetilde\psi_R,\psi_R)$, and \emph{odd},
$(\Phi,\lambda_2)$ and $(\widetilde\psi_L,\psi_L)$, $N=1$ 
superfields~\footnote{Of course the above lifting on hypermultiplets
is arbitrary and we could equally well consider 
$(\widetilde\psi_R,\psi_R)$ as odd
and  $(\widetilde\psi_L,\psi_L)$ as even.}.
After the $\mathbb{Z}_2$ projection the only
surviving zero modes are in $N=1$ vector and chiral multiplets. If the
chiral multiplet 
$(\widetilde\psi_R,\psi_R)$ is not in a real representation of the
gauge group, a multiplet of opposite chirality localized in the 4D
boundary $(\widetilde\psi_L,\psi_L)$ can be introduced 
to cancel anomalies. In
order to do that $\mathbb{Z}_2$ must have a further action on
the boundary under which all boundary states are 
odd~\cite{dq2}. 
This action can be defined as $(-1)^{\varepsilon_i}$ for the
chiral multiplet $X_i$, such that $\varepsilon_i=1$ (0) for $X_i$
living in the brane (bulk), which creates a selection rule for 
superpotential interactions on the brane. 

The MSSM is then made up of zero
modes of fields living in the 5D
bulk and chiral $N=1$ multiplets in the 4D brane, at 
localized points of the bulk. A superpotential interaction can only exist at
the brane of the type $UTT$ or $UUU$ to satisfy the orbifold selection
rule~\cite{sharpe}, 
although the latter, $UUU$, are expected to have Yukawa couplings which
are suppressed with respect to those in $UTT$ by a factor 
$(R\Lambda_s)^{-1}$, and corresponding to the fact that 
localized couplings of states propagating in the bulk must
be (volume) suppressed.

To allow for the MSSM superpotential on the brane,
\begin{equation}
\label{superpotential}
W=\left[h_U\, Q\,H_2\,U+h_D\, Q\,H_1\,D+h_E\, L\,H_1\,E+\mu\, H_2\,H_1
\right]\delta(x_5)
\end{equation}
and to be consistent with the orbifold selection rule and with a top/stop 
hypermultiplet propagating in the bulk, to trigger a finite electroweak
radiative breaking, the only solution is that the $SU(2)_L$ singlets,
$\mathbb{U},\, \mathbb{D},\, \mathbb{E}$ propagate in the 
bulk~\footnote{Here we assume that all three
generations propagate in the same way. Otherwise they can produce an accute
flavor problem and trigger strong $CP$ violation, which translate into
strong bounds on the scale $1/R$~\cite{alex2}.}. In this case the $SU(2)_L$
doublets $Q,\, L,\, H_2,\ H_1$ are localized on the brane and transform as
chiral $N=1$ multiplets~\footnote{There is technically speaking another
possibility: that matter doublets $\mathbb{Q},\, \mathbb{L}$ 
are living in the bulk and 
matter singlets and Higgs doublets $U,\, D,\, E,\, H_2,\ H_1 $ are
localized on the brane. This possibility looks less natural and will not
be explicitly considered in this paper, although it leads to 
results very similar to those that will be found. It also leads to
a drawback in the supersymmetric spectrum concerning the lightest
supersymmetric particle, as we will comment later on.}. 

Since both Higgs fields are on the brane, the $\mu$-parameter does not arise
through compactification~\cite{alex1} and, although it is an allowed term in
the superpotential, one has to consider it as an effective parameter and 
rely on its generation as a result of the integration of the massive states of
the underlying (supergravity or string) 
theory~\footnote{Another possibility, that has been
recently pointed out in Ref.~\cite{Hall}, is having a singlet field $S$ in
the bulk acquiring a vacuum expectation value (VEV) by radiative corrections 
induced by another field, in a similar way 
to the one by which the top/stop sector makes the 
Higgs field to acquire a VEV in this paper. Of course this
situation requires enlarging the MSSM to the NMSSM and the Higgs sector
gets mixed with the singlet states~\cite{singlet}.}. 
In this sense the situation is no better than
in the usual MSSM, except for the fact that the cutoff of the theory, 
$\Lambda_s$, the
scale at which the structure of the underlying theory should be considered,
is at most two orders of magnitude larger that the scale of
supersymmetry breaking and therefore a modest suppression should be sufficient
for phenomenological purposes. 

After compactification on $S^1/\mathbb{Z}_2$ the zero mode of the vector 
multiplet $\mathbb{V}^{\,(0)}$ is just the 4D MSSM $N=1$ vector multiplet
$(A^{(0)}_\mu,\lambda^{(0)}_1)$, while the massive modes are
the massive 4D $N=2$ vector multiplets
$(A^{(n)}_\mu,\lambda_1^{(n)},\lambda_2^{(n)},\Sigma^{(n)})$, 
with a mass $n/R$, where
$\lambda_1^{(n)},\ \lambda_2^{(n)}$ are Majorana spinors.
Similarly the zero mode of matter 
hypermultiplets $\mathbb{H}^{(0)}$ is the 4D $N=1$ chiral multiplet
$(\widetilde\psi^{(0)}_R,\psi^{(0)}_R)$, while the massive modes are
4D $N=2$ hypermultiplets, 
$(\widetilde\psi^{(n)}_R,\widetilde\psi^{(n)}_L,\psi^{(n)})$, with a mass 
$n/R$, where $\psi^{(n)}$ is a Dirac spinor with components
$(\psi_R^{(n)},\psi_L^{(n)})$. Finally, the chiral fields which are
localized on the brane are massless, except for the supersymmetric mass term
$\mu$ introduced in the superpotential (\ref{superpotential}) that gives
a common mass to Higgs bosons and higgsinos.

Supersymmetry breaking was performed in Refs.~\cite{PQ,savas,alex1} using
the SS mechanism based on the subgroup $U(1)_R$ which survives after the
orbifold action $S^1/\mathbb{Z}_2$. Using the $R$-symmetry $U(1)_R$ with
parameter $\omega$ to impose different boundary conditions for bosons and
fermions inside the 5D $N=1$ multiplets, one obtains for the $n$-th 
Kaluza-Klein (KK) mode of gauge bosons and chiral fermions living in the bulk
the compactification mass $n/R$, while for the $n$-th mode of gauginos,
$\lambda^{(n)}$, and supersymmetric partners of chiral fermions living in the 
bulk, $\widetilde\psi_R^{(n)},\widetilde\psi_L^{(n)}$ the mass
$(n+\omega)/R$. Notice that for the particular case $\omega=1/2$ the gauginos
$\lambda^{(n)}$ and $\lambda^{-(n+1)}$, $n>0$, are degenerate in mass
and constitute a Dirac fermion. A detailed discussion was done in 
Ref.~\cite{PQ} for the case of gauginos. For matter scalars in hypermultiplets,
$(\widetilde{\psi}_R,\widetilde{\psi}_L)$, that transform under the subgroup
$U(1)_R$ we can write the SS boundary conditions as,
\begin{equation}
\label{SScond}
\left(
\begin{array}{c}
\widetilde{\psi}_R \\
\widetilde{\psi}_L
\end{array}
\right)=
\left[
\begin{array}{cc}
\cos \omega x_5/R & -\sin \omega x_5/R \\
\sin \omega x_5/R & \cos \omega x_5/R
\end{array}
\right]
\left(
\begin{array}{c}
\varphi_R \\
\varphi_L
\end{array}
\right)
\end{equation}
where $\varphi_{R}$ ($\varphi_{L}$) are even (odd) periodic functions  
$\varphi_{R,L}(x_5)=\varphi_{R,L}(x_5+2\pi R)$. Making a Fourier expansion
along the $x_5$ direction with coefficients $\varphi_{R,L}^{(n)}$ we can write:
\begin{align}
\widetilde{\psi}_R=& \sum_{n=-\infty}^{\infty} \cos\frac{(\omega+n)\,x_5}{R}\
\widetilde{\psi}_R^{(n)}\nonumber\\
 \widetilde{\psi}_L=& \sum_{n=-\infty}^{\infty} \sin\frac{(\omega+n)\,x_5}{R}\
\widetilde{\psi}_L^{(n)}
\label{expansion}
\end{align}
where 
\begin{align}
\widetilde{\psi}_R^{(n)}& \equiv \widetilde{\psi}_L^{(-n)}=
\frac{1}{2} \left( \varphi_R^{(n)}- \varphi_L^{(n)}
\right),\quad n\geq 0
\nonumber\\
\widetilde{\psi}_L^{(n)}& \equiv \widetilde{\psi}_R^{(-n)}=
\frac{1}{2} \left( \varphi_R^{(n)}+ \varphi_L^{(n)}
\right),\quad n\geq 0
\label{modos}
\end{align}
are the mass eigenstates modes.

In this way the MSSM states, made out of bulk zero modes and localized states,
acquire tree-level masses: gauge bosons, right-handed fermions
and left-handed chiral multiplets localized on the brane
are massless; gauginos and 
right-handed sfermions are massive, with masses $\omega/R$; 
Higgs bosons and 
higgsinos, localized on the brane, are massive with a common
supersymmetric mass $\mu$.

Since the Higgs bosons have a positive squared mass $\mu^2$, electroweak
symmetry is preserved at the tree-level. However,
we will see in the next section how this 
tree-level mass  pattern, along with the Yukawa
couplings contained in the superpotential (\ref{superpotential}), and in
particular the top Yukawa coupling $h_t$, will be able to break radiatively 
the electroweak symmetry and provide a well defined spectrum for the Higgs
masses and the masses of the supersymmetric partners localized on the brane.

\section{Radiative electroweak breaking}  

The Higgs potential along the direction of the neutral components of
the fields $H_2=h_2+i\, \chi_2$ and $H_1=h_1+i\, \chi_1$ can be written as,
\begin{eqnarray}
\label{potencial}
V(H_1,H_2)&=& m_1^2\,|H_1|^2+m_2^2\,|H_2|^2+
m_3^2\, \left(H_1\cdot H_2+h.c.\right) +\frac{g^2+g^{\prime 2}}{8}
\left(|H_1|^2-|H_2|^2\right)^2\nonumber\\
&+&\lambda_t\, |H_2|^4+\lambda_b\, |H_1|^4
\end{eqnarray}
where the supersymmetric tree-level relations $m_1=m_2=\mu$ and $m_3^2=0$,
$\lambda_t=\lambda_b=0$ hold. These relations are spoiled by radiative 
corrections which provide contributions to all the above parameters. These
corrections are driven by the $SU(2)_L\times U(1)_Y$ gauge couplings 
$g$ and $g^\prime$, and by the top and bottom Yukawa couplings, defined as:
\begin{equation}
\label{Yukawa}
h_t=\frac{m_t}{v}\, \sqrt{\frac{1+t^2_\beta}{t^2_\beta}},\quad
h_b=\frac{m_b}{v}\, \sqrt{1+t^2_\beta}
\end{equation}
where $t_\beta\equiv\tan\beta\equiv v_2/v_1$, $v_i=\langle H_i \rangle$
are the vacuum expectation values of the Higgs fields,
$v=\sqrt{v_1^2+v_2^2}=174.1$ GeV, and $m_t$ and $m_b$ are the top and bottom
running masses. Notice that $h_b$ can become important only for large values
of $t_\beta$, as those that will be found by minimization of the one-loop
effective potential. We will consider the leading radiative corrections:
in particular $g^2$-corrections to the
quadratic terms which are zero at the tree-level ($m_3^2$) and $h_{t,b}^4$
corrections to the quartic terms which are $\mathcal{O}(g^2)$ at the 
tree-level. For this reason we will not consider any radiative term as
$\left(H_1\cdot H_2\right)^2$ in Eq.~(\ref{potencial}) and neglect 
$g^{\prime 2}$-radiative corrections in the numerical analysis.

All radiative corrections to the potential parameters in (\ref{potencial})
will depend on $1/R$ and $\omega$. In particular the one-loop radiative
corrections to any scalar localized on the brane were computed in 
Ref.~\cite{alex1} where the corresponding diagrams were 
identified~\footnote{See Figs. 2 and 3 of Ref.~\cite{alex1}.}.
A simple application to the Higgs mass terms $m_1^2$ and $m_2^2$ yields:
\begin{eqnarray}
\label{masasr}
m_2^2&=&\mu^2-\frac{6 h_t^2-3 g^2}{32 \pi^4}\frac{\Delta(\omega)}{R^2}
\nonumber\\ 
m_1^2&=&\mu^2-\frac{6 h_b^2-3 g^2}{32 \pi^4}\frac{\Delta(\omega)}{R^2}
\end{eqnarray}
where the function $\Delta$ is defined as
\begin{equation}
\label{Delta}
\Delta(\omega)=2 \zeta(3)-\left[Li_3(r)+Li_3(1/r)\right]\ ,
\end{equation}
$r=\exp\left(2\pi i\omega\right)$, and $Li_n(x)=\sum_{k=1}^\infty x^k/k^n$ is
the polylogarithm function of order $n$.

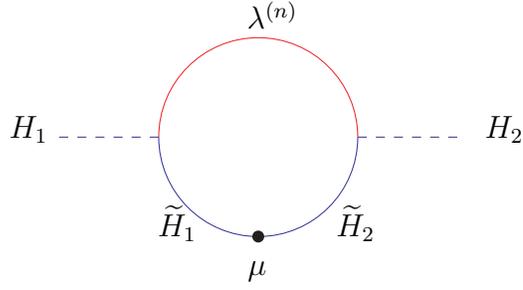
\begin{figure}[H]
\begin{center}
\setlength{\unitlength}{.75pt}
\SetScale{0.75}
\begin{picture}(300,100)(0,0)
\thicklines
\put(25,50){$H_1$}
\put(265,50){$H_2$}
\put(145,105){$\lambda^{(n)}$}
\put(100,0){$\widetilde{H}_1$}
\put(190,0){$\widetilde{H}_2$}
\SetColor{Red}
\CArc(150,50)(50,0,180)
\SetColor{Blue}
\CArc(150,50)(50,180,360)
\SetColor{Blue}
\DashLine(50,50)(100,50){5}  %\Vertex(110,50){2}
\DashLine(200,50)(250,50){5} %\Vertex(190,50){2}
\SetColor{Black}
\Vertex(150,0){3}
%\Text(140,0)[1]{mu}
\end{picture}  \\ { $\mu$} 
\end{center}
\caption{One-loop diagrams contributing to $m_3^2$.}
\label{Fig:m32}
\end{figure}

The mass term $m_3^2$ in (\ref{potencial}) is generated by the one-loop diagram
exchanging KK-modes of gauginos, $\lambda^{(n)}$ and localized higgsinos,
$\widetilde{H}_{1,2}$, as shown in Fig.~\ref{Fig:m32}.
\vspace{1cm}
\noindent
The resulting contribution is given by,
\begin{equation}
\label{m32}
m_3^2=\mu \frac{3\, g^2}{512\pi^2} \frac{\omega}{R}\left[
i\, Li_2(r)-i\,  Li_2(1/r)\right]
\end{equation}
Notice that, for the particular case $\omega=1/2$ ($r=-1$), $m_3^2=0$,
reflecting the fact that the gauginos $\lambda^{(n)}$ are, in that
case, Dirac fermions, as it was already mentioned.

The quartic couplings $\lambda_{t,b}$ are generated at the one-loop level 
by loop diagrams exchanging KK-modes, $\widetilde{t}^{(n)}_R$ and 
$\widetilde{b}^{(n)}_R$, and localized modes, $\widetilde{t}_L$ and 
$\widetilde{b}_L$. The diagrams contributing to $\lambda_t$ are shown in 
Fig.~\ref{Fig:lambdat}. The contribution to $\lambda_b$ being similar, 
just changing $H_2\to H_1$ and $t\to b$.
\vspace{1.cm}
\begin{figure}[H]
\setlength{\unitlength}{.7pt}
\SetScale{0.7}
\begin{picture}(300,100)(0,0)
\thicklines
\put(25,95){$H_2$}
\put(25,5){$H_2$}
\put(265,95){$H_2$}
\put(265,5){$H_2$}
\put(130,110){$\widetilde{t}_R^{(n)},\widetilde{t}_L$}
\put(130,-20){$\widetilde{t}_R^{(m)},\widetilde{t}_L$}
\SetColor{PineGreen}
\DashCArc(150,50)(50,0,180){5}
\DashCArc(150,50)(50,180,360){5}
\SetColor{Blue}
\DashLine(50,95)(100,50){5}  
\DashLine(50,5)(100,50){5}
\DashLine(200,50)(250,95){5} 
\DashLine(200,50)(250,5){5}
\end{picture}  
\setlength{\unitlength}{.7pt}
\SetScale{0.7}
\begin{picture}(300,100)(0,0)
\thicklines
\put(25,110){$H_2$}
\put(25,-10){$H_2$}
\put(265,95){$H_2$}
\put(265,5){$H_2$}
\put(145,110){$\widetilde{t}_L$}
\put(145,-20){$\widetilde{t}_L$}
\put(75,50){$\widetilde{t}_L^{\,(n)}$}
\SetColor{Blue}
\DashCArc(150,50)(50,0,135){5}
\SetColor{Red}
\DashCArc(150,50)(50,135,225){5}
\SetColor{Blue}
\DashCArc(150,50)(50,225,360){5}
\SetColor{Blue}
\DashLine(50,110)(114.64,85.36){5}  
\DashLine(50,-10)(114.64,14.64){5}
\DashLine(200,50)(250,95){5} 
\DashLine(200,50)(250,5){5}
\end{picture}  
\end{figure}
\vspace{.5cm}
\begin{figure}[H]
\setlength{\unitlength}{.7pt}
\SetScale{0.7}
\begin{picture}(300,100)(0,0)
\thicklines
\put(25,110){$H_2$}
\put(25,-10){$H_2$}
\put(265,110){$H_2$}
\put(265,-10){$H_2$}
\put(145,110){$t_L$}
\put(145,-15){$t_L$}
\put(75,50){$t_R^{\,(n)}$}
\put(215,50){$t_R^{\,(m)}$}
\SetColor{Red}
\CArc(150,50)(50,-45,45)
\SetColor{Blue}
\CArc(150,50)(50,45,135)
\SetColor{Red}
\CArc(150,50)(50,135,225)
\SetColor{Blue}
\CArc(150,50)(50,225,315)
\SetColor{Blue}
\DashLine(50,110)(114.64,85.36){5}  
\DashLine(50,-10)(114.64,14.64){5}
\DashLine(185.36,85.36)(250,110){5} 
\DashLine(185.36,14.64)(250,-10){5}
\end{picture}  
\setlength{\unitlength}{.7pt}
\SetScale{0.7}
\begin{picture}(300,100)(0,0)
\thicklines
\put(25,110){$H_2$}
\put(25,-10){$H_2$}
\put(265,110){$H_2$}
\put(265,-10){$H_2$}
\put(145,110){$\widetilde{t}_L$}
\put(145,-20){$\widetilde{t}_L$}
\put(75,50){$\widetilde{t}_L^{\,(n)}$}
\put(215,50){$\widetilde{t}_L^{\,(m)}$}
\SetColor{Red}
\DashCArc(150,50)(50,-45,45){5}
\SetColor{Blue}
\DashCArc(150,50)(50,45,135){5}
\SetColor{Red}
\DashCArc(150,50)(50,135,225){5}
\SetColor{Blue}
\DashCArc(150,50)(50,225,315){5}
\SetColor{Blue}
\DashLine(50,110)(114.64,85.36){5}  
\DashLine(50,-10)(114.64,14.64){5}
\DashLine(185.36,85.36)(250,110){5} 
\DashLine(185.36,14.64)(250,-10){5}
\end{picture}  
\vspace{1cm}
\caption{One-loop diagrams contributing to $\lambda_t$.}
\label{Fig:lambdat}
\end{figure}
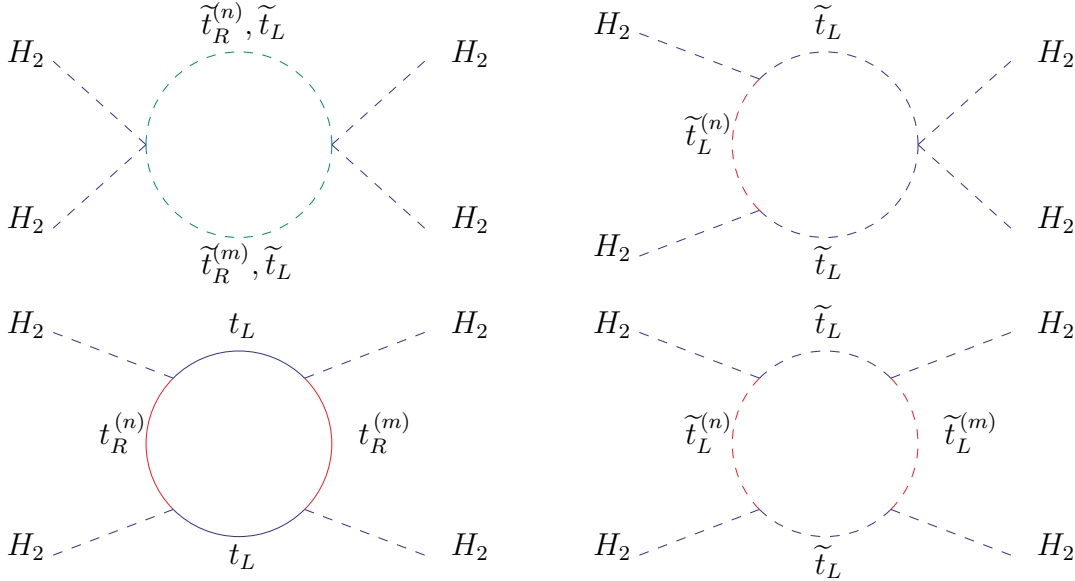

Notice that in Fig.~\ref{Fig:lambdat} we are sometimes propagating the
modes $\widetilde{t}_L^{(n)}$. They correspond, in our notation, to the
odd modes of the hypermultiplets $(\widetilde{t}_R^{(n)},t_R^{(n)};
\widetilde{t}_L^{(n)},t_L^{(n)})$ that can couple to the brane through
$\partial_5$ couplings, and should not be confused with the localized
multiplets $(\widetilde{t}_L,t_L)$. The resulting expression is,
\begin{eqnarray}
\label{lambdas}
\lambda_t&=&-\frac{3 h_t^4}{8\pi^2}\left\{ -1+\log 2\pi R M_Z\right.
\\
&-&\left.
\frac{1}{4\left(r^2-1\right)}\left[(r^2-1)\left(\log(1-r)+\log(1-1/r)\right)
+(1+r^2)\left(Li_2(1/r)-Li_2(r)\right)\right]\right\}\nonumber
\end{eqnarray}
and a similar expression for $\lambda_b$ just changing $h_t\to h_b$

\begin{figure}[H]
%\centering
\epsfig{file=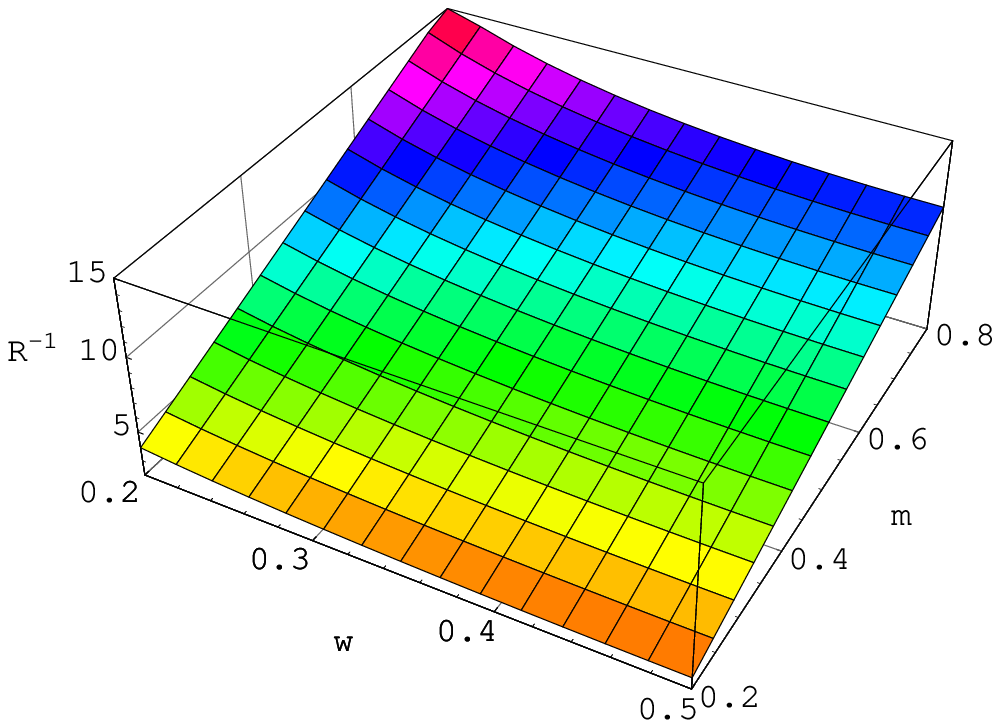,width=0.48\linewidth}\hspace{.3cm}
\epsfig{file=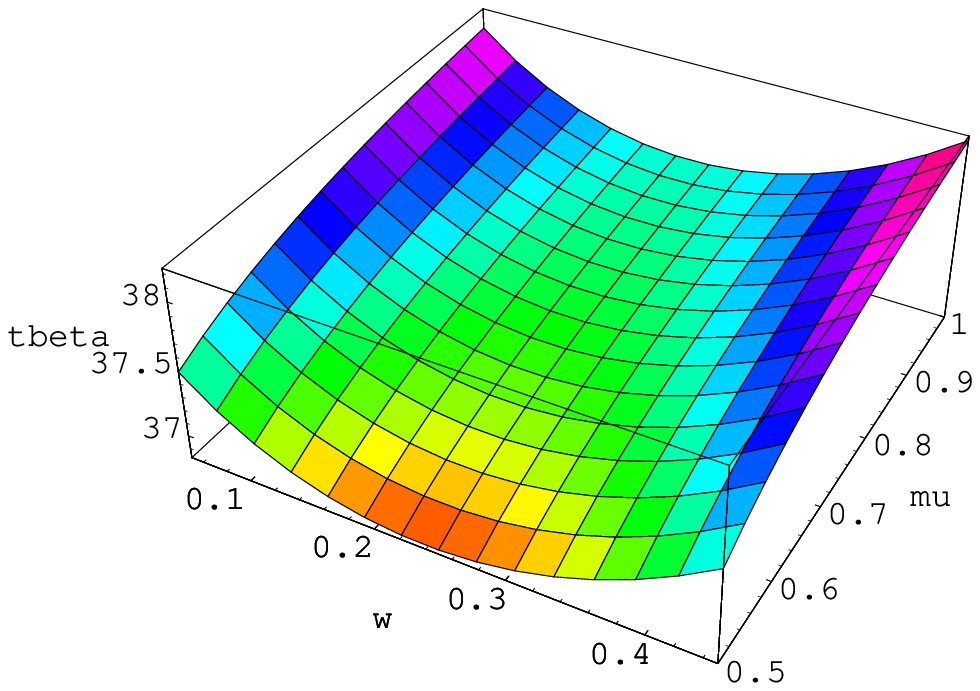,width=0.48\linewidth}
\caption{Plots of $1/R$, in TeV (left panel) and $t_\beta$ (right panel), 
as functions of $\omega$ and $\mu$, in TeV, as given from the minimization
conditions (\ref{minimo}).}     
\label{Fig:Rtbeta}
\end{figure}

Finally we have as free parameters,
$1/R$, $\omega$, $\mu$ and $t_\beta$. Two of them will be fixed by the
minimization conditions $V'_{h_1}=V'_{h_2}=0$ which read as
\begin{align}
\label{minimo}
3\left(1+\frac{1}{t_\beta^2}\right)\frac{2\,h_t^2-g^2}{32\,\pi^4}
\frac{\Delta(\omega)}{R^2}&=
\mu^2+\frac{1}{2}M_Z^2+2\lambda_t v^2
+ \left(\mu^2-m_A^2-\frac{1}{2}M_Z^2\right)\frac{1}{t_\beta^2}\nonumber\\ & \\
\frac{2h_t^2-g^2}{2h_b^2-g^2}&= 
\frac{\mu^2+\frac{1}{2}M_Z^2+2\lambda_t v^2
+ \left(\mu^2-m_A^2-\frac{1}{2}M_Z^2\right)/t_\beta^2}
{\mu^2-m_A^2-\frac{1}{2}M_Z^2+\left(\mu^2-\frac{1}{2}M_Z^2
+2\lambda_b v^2\right)/t_\beta^2}\ ,\nonumber
\end{align}
where
\begin{equation}
\label{mA}
m_A^2=-m_3^2\ \frac{1+t_\beta^2}{t_\beta}
\end{equation}
is the mass of the CP-odd Higgs once the minimization conditions (\ref{minimo})
have been used. In fact we have chosen to select $1/R$ and $t_\beta$ as 
functions of the other variables. The corresponding plots are shown in
Fig.~\ref{Fig:Rtbeta}.
We can see from Fig.~\ref{Fig:Rtbeta} that the minimization conditions impose
a solution with large $\tan\beta$ ($t_\beta\simeq 35-40$) and values of the
compactification scale going from a few TeV to $\sim 10-15$ TeV, 
depending on the
values of $\omega$ and $\mu$. 

SM precision measurements settle bounds on
electroweak observables which, for higher dimensional models with
gauge fields living in the bulk of the extra dimension, 
translate on lower bounds on
the compactification scale $1/R$. The model we are studing was analyzed in
Ref.~\cite{alex2}, where a very general class of models was considered. 
We found
that, for large values of $t_\beta$, the lower bound on $1/R$ is $\sim 4$ 
TeV which is in the ballpark provided by Fig.~\ref{Fig:Rtbeta}.

\section{The Higgs mass spectrum}

\begin{figure}[H]
%\centering
\epsfig{file=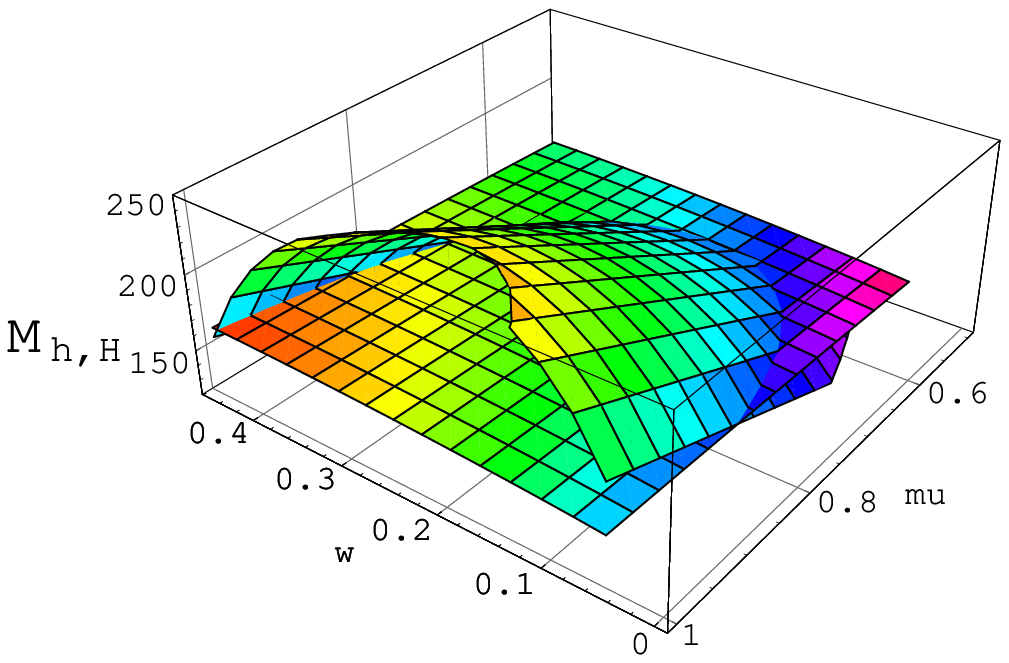,width=0.48\linewidth}\hspace{.3cm}
\epsfig{file=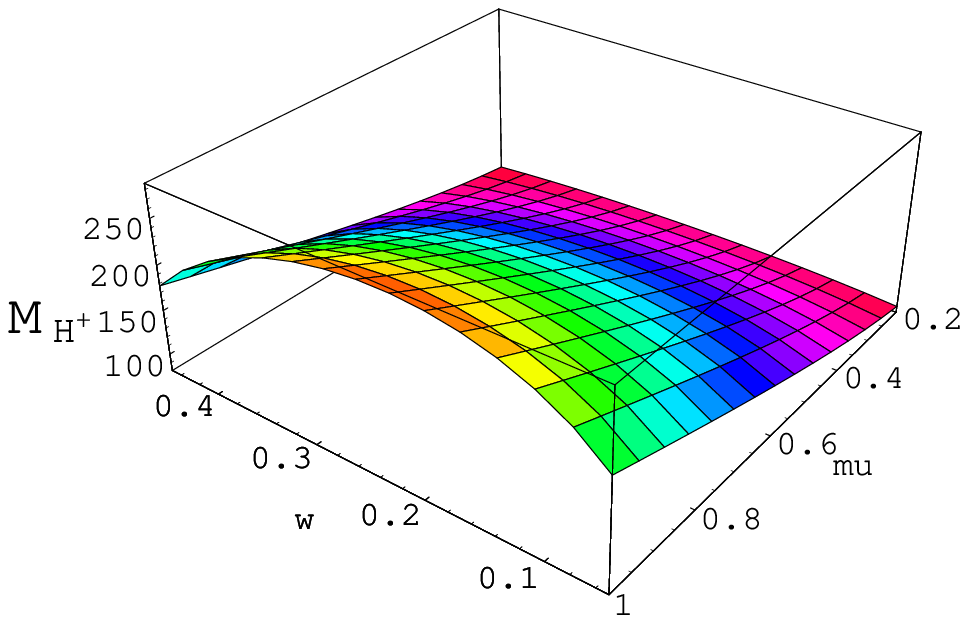,width=0.48\linewidth}
\caption{Plots of $M_{h,H}$, in GeV (left panel) and $M_{H^{\pm}}$, in GeV
(right panel), as functions of $\omega$ and $\mu$, in TeV.}     
\label{Fig:masas}
\end{figure}

The neutral Higgs sector has one CP-odd and two CP-even scalar bosons.
The mass of the  CP-odd Higgs boson was already given in Eq.~(\ref{mA}).
The squared mass matrix for neutral CP-even scalar bosons is given by,
\begin{equation}
\label{masahH}
\mathcal{M}_0^2=\left[
\begin{array}{cc}
m_A^2 s^2_\beta+\left( M_Z^2+4 \lambda_b\, v^2 \right)c^2_\beta &
\left(m_A^2+M_Z^2\right) s_\beta\, c_\beta \\ & \\
\left(m_A^2+M_Z^2\right) s_\beta\, c_\beta & 
m_A^2 c^2_\beta+\left( M_Z^2+4 \lambda_t\, v^2 \right)s^2_\beta 
\end{array}
\right]
\end{equation}
In the large $t_\beta$-limit the two eigenvalues are:
\begin{equation}
M_H^2=m_A^2,\quad
M_h^2=M_Z^2+4\, \lambda_t\, v^2\ ,
\label{massainf}
\end{equation}
where $h$ is the SM-like Higgs and $H$ the Higgs with non-SM couplings,
while the mass of the charged Higgs $H^{\pm}$ is given by
\begin{equation}
\label{mch}
M_{H^{\pm}}^2=m_A^2+\, M_W^2
\end{equation}
The masses of the neutral CP-even and charged Higgs bosons are shown in
Fig.~\ref{Fig:masas}.
In the left panel of Fig.~\ref{Fig:masas}, $M_h$ is the flat surface while
$M_H$ corresponds to the steepest one. The crossing is characteristics of
the large $t_\beta$ solution. For large values of $\mu$, $M_h$ is the
lightest (SM-like) Higgs mass. Notice that the mass $M_h$ is not controlled
by the compatification scale $1/R$, but only by the weak scale $v$. This is
a reflection of a similar behaviour in the MSSM where the lightest SM-like
Higgs mass is not controlled by the supersymmetry breaking scale. On the other
hand, for small values of $\mu$, $M_H$ corresponds to the lightest Higgs mass.
Its mass is controlled by the compactification scale.

To make contact with Refs.~\cite{Barbieri,Hall} we can fix $\omega=1/2$. In
that case the $m_3^2$-term is not generated, as we said previously, and 
$m_A=0$, unless we introduce an additional term like 
$\sim(\lambda/\Lambda_s)\, \left( H_1 H_2\right)^2$ in the superpotential, 
giving rise to
an $m_3^2$-term as $\sim \lambda\mu v^2/\Lambda_s$. In that case the Higgs
spectrum depends on the parameters $\mu$ and $\lambda$, with some sensitivity
on the cutoff $\Lambda_s$. We recover the results of 
Ref.~\cite{Barbieri} in the limit $\lambda,\mu\to 0$, in which case we
obtain $M_h\simeq 128$ GeV, in agreement with the result in \cite{Barbieri}.
However, in general, for $\omega\neq n/2$ the PQ invariance is broken by the
gaugino masses and there is no need for the $\lambda$-term in the 
superpotential. 

Finally we will comment on the constraints imposed on our model from the
Higgs searches at LEP. Preliminary results of last year run show a lower
limit on the SM Higgs mass about 113 GeV. On the other hand, 
an excess of candidates for the process $e^+ e^-\to Z^*
\to Zh$ has been reported by the ALEPH  and L3 Collaborations 
for center-of-mass energies $\sqrt{s}>206$ GeV, for a SM-Higgs with a
mass around $M_h\simeq 115$ GeV~\cite{Higgsex}, which decays predominantly
into $b\bar{b}$. Now the question is whether this mass can be accomodated in 
our model.
A quick glance at Fig.~\ref{Fig:masas} (left panel) shows that such low
masses should be described by what we name as $H$ eigenstate, with a mass
$m_H\simeq m_A$. In the large $t_\beta$ region, the state $H$ is predominantly
$H_1^0$, with unconventional couplings to gauge bosons and fermions, as
opposed to the state $h$, the SM-like Higgs boson, with SM-like couplings to
gauge bosons and fermions, and 
with a mass entirely controlled by the electroweak
breaking parameter. However in this region we have $M_h>M_H$. In this way the 
coupling $ZZH_1=(ZZh)^{SM}c_\beta$ is strongly suppressed as 
$\sim 1/t_\beta$ and so does the $H$ direct production. In other words
a SM-like Higgs with a mass $\sim$ 115 GeV cannot be accomodated by our 
present model~\footnote{We thank J.R.~Espinosa and
C.~Wagner for pointing out this to us.}.

\begin{figure}[htb]
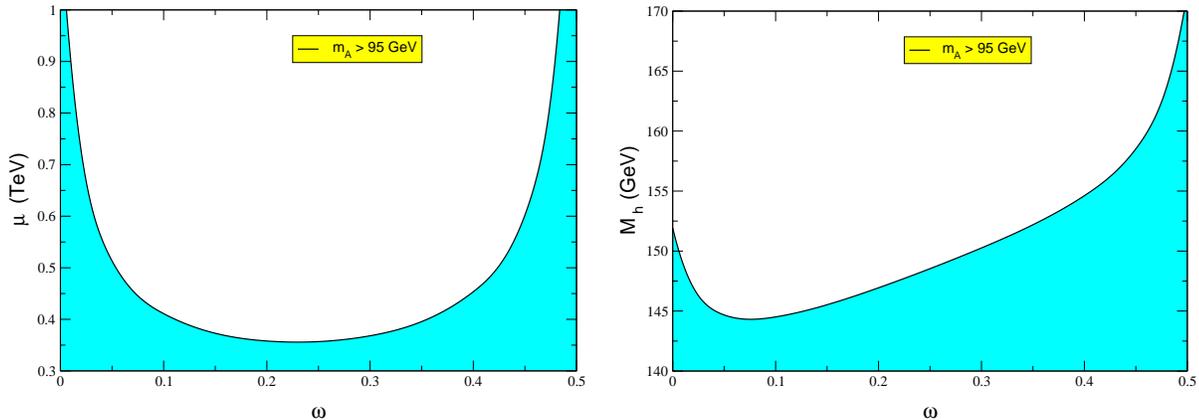

%\centering
\vspace{5mm}
\epsfig{file=muomega.eps,width=0.48\linewidth}\hspace{.3cm}
\epsfig{file=mhomega.eps,width=0.48\linewidth}
\caption{Lower limit on $\mu$ from the 
LEP bound $m_A\simgt 95$ GeV (left panel), and the corresponding 
bound for $M_h$ (right panel).}     
\label{Fig:LEP}
\end{figure}

What are then the limits imposed to our model by the bounds on Higgs searches
at LEP? For large $t_\beta$ and moderate values of the pseudoscalar mass
$m_A$, such that $m_A=M_H$, LEP searches on the MSSM Higgs sector,
based on the process 
$e^+ e^-\to H A$, settle a lower bound on $m_A$ as $m_A \simgt 95$ GeV.
This constraint translates into a lower limit $\mu\simgt \mu(\omega)$ on the
$\mu$-parameter that is shown in Fig.~\ref{Fig:LEP} (left panel) where the
shadowed region is excluded. In particular 
an absolute lower bound on $\mu$ around 350 GeV can be read off the plot.
The corresponding lower bound on the
SM-like Higgs mass, $M_h$, is shown in Fig.~\ref{Fig:LEP} (right panel) 
where again the shadowed region is the excluded one. From this plot we can
see that the absolute lower limit for the SM-Higgs mass (in the one-loop
approximation) is $\sim$ 145 GeV. This mass is probably too heavy to be
discovered at Tevatron and should await till the LHC collider.

In summary we see that the Higgs sector of this model is very constrained
and will be probed in the next generation of colliders 
(LHC). It predicts, as any MSSM with large $t_\beta$, an
almost degeneracy between the masses of one of the CP-even and the CP-odd
states. LEP bounds on the MSSM Higgs sector set an absolute lower bound on 
the SM-like Higgs mass, around 145 GeV, and the $\mu$-parameter, around
350 GeV, which provides the higgsino masses. The model is very predictive 
and will be fully tested at LHC. 

\section{The supersymmetric spectrum}

The supersymmetric mass spectrum is fully determined by the mechanism of
electroweak and supersymmetry breaking that we have described in previous
sections. All soft breaking mass terms of zero modes propagating in the
bulk of the fifth dimension are equal, and given by the SS supersymmetry
breaking mechanism,
\begin{equation}
\label{softbulk}
M_1=M_2=M_3=m_{U_i}=m_{D_i}=m_{E_i}=A_{abc}=\frac{\omega}{R}
\end{equation}
where $i=1,2,3$ labels the generation number and 
$A_{abc}$ denotes generically all
soft couplings corresponding to trilinear terms in the superpotential. For 
example, in the case of the top quark coupling in the superpotential,
$h_t\, H_2\, Q\, U_R$, $A_t$ comes from the term in the 5D Lagrangian
\begin{equation}
\label{5DA}
\mathcal{L}_5=-h_t\,H_2\, \widetilde{Q}\, \left(\partial_5\, 
\widetilde{U}_L\right)\, \delta(x_5)+h.c.
\end{equation}
giving rise, after dimensional reduction and SS breaking, to the 4D Lagrangian
\begin{equation}
\label{4DA}
\mathcal{L}_4=-\sum_{n=-\infty}^{\infty} 
h_t\,\frac{n+\omega}{R}\,H_2\, \widetilde{Q}\, \widetilde{U}^{(n)}_L+h.c.
\end{equation}
in the notation of Eqs.~(\ref{expansion}) and (\ref{modos}).
The term $n=0$ of (\ref{4DA}) gives rise to $A_t$. 

Concerning now the $N=1$ sector localized on the brane, it does not receive
any tree-level mass, except for the higgsinos 
(and Higgs bosons) that get a tree-level mass
$\mu$. However the scalar supersymmetric partners of left-handed quarks and
leptons do receive a radiative mass from the bulk fields, where supersymmetry
is broken, mediated by gauge and Yukawa interactions. These contributions
were computed in full generality in Ref.~\cite{alex1} and we will just quote
here the corresponding result applied to the present model.

Squarks receive the main radiative contribution from the gluon/gluino
sector proportional to the QCD gauge coupling $g_3$. For the first and
second generation squarks $\tilde{q}$ we can neglect all other gauge and
Yukawa couplings and write,
\begin{equation}
\label{squarks12}
m^2_{\tilde{q}}\simeq \frac{8}{9}\, \frac{g_3^2}{h_t^2-g^2/2} \left(
\mu^2+\frac{1}{2}M_Z^2\right)
\end{equation}
Plugging numbers in Eq.~(\ref{squarks12}) we can see that it yields
$m_{\tilde{q}}\simeq 1.4\,\mu$.

For the third generation squark doublet 
$\widetilde{Q}=(\tilde{t}_L,\tilde{b}_L)$ there is,
on top of the gauge contribution of (\ref{squarks12}), a negative contribution
from $h_{t,b}$ Yukawa couplings. Using again the minimization conditions we
can write,
\begin{equation}
\label{squarks3}
m_{\widetilde{Q}}^2\simeq \left(1-\frac{3(h_t^2+h_b^2)}
{8\, g_3^2}\right)\, m_{\tilde{q}}^2
\end{equation}
which gives the numerical rough estimate $m_{\tilde{Q}}
\simeq \mu$.

The mixing can be neglected for all states except for the third generation
of up-type squarks for which the two mass eigenvalues can be approximately
written as,
\begin{align}
\label{stops}
M_{\tilde{t}_2}^2 \simeq & \left(\frac{\omega}{R}\right)^2+2\, m_t^2
+\frac{m_{\widetilde{Q}}^2}{\left(\frac{\omega}{R}\right)^2 
-m_{\widetilde{Q}}^2}\, m_t^2\nonumber\\
M_{\tilde{t}_1}^2 \simeq &\ m_{\widetilde{Q}}^2-
\frac{m_{\widetilde{Q}}^2}{\left(\frac{\omega}{R}\right)^2 
-m_{\widetilde{Q}}^2}\, m_t^2
\end{align}

For the supersymmetric partners of lepton doublets 
$\widetilde{L}=(\tilde{\nu}_L,\tilde{\ell}_L)$ (the three generations) we
obtain, in the same way, the soft squared mass value,
\begin{equation}
\label{slepton}
m_{\tilde{L}}^2\simeq \,
\frac{9\, \alpha_2}{16\,\alpha_3}\, m_{\tilde{q}}^2
\end{equation}
which gives the numerical estimate $m_{\tilde{L}} \simeq 0.55\,\mu$.
In this way, and neglecting the mixing, the mass of charged sleptons can
be approximated by,
\begin{align}
\label{sleptons}
M_{\tilde{\ell}_L}^2\simeq &\ m_{\tilde{L}}^2+m_{\ell}^2
+M_Z^2\left(\frac{1}{2}-s^2_W\right)\ \frac{t_\beta^2-1}{t_\beta^2+1}
\nonumber\\
M_{\tilde{\ell}_R}^2\simeq &\, \left(\frac{\omega}{R}\right)^2+m_{\ell}^2
+M_Z^2\, s^2_W\, \frac{t_\beta^2-1}{t_\beta^2+1}
\end{align}
while the mass of the sneutrinos 
\begin{equation}
\label{sneutrino}
M_{\tilde{\nu}_L}^2\simeq\, m_{\tilde{L}}^2-\frac{1}{2}M_Z^2\,
\frac{t_\beta^2-1}{t_\beta^2+1}
\end{equation}
In this way the $\tilde{\nu}_L$ turns out to be the lightest supersymmetric
particle (LSP) while the slepton $\tilde{\ell}_L$ is the next to lightest
supersymmetric particle (NLSP). Their masses satisfy the approximate
relation
\begin{equation}
\label{approx}
M_{\tilde{\ell}_L}^2-M_{\tilde{\nu}_L}^2\simeq\, M_Z^2
\end{equation}
Since s-neutrinos are neutral particles any kind of cosmological problems
associated with the existence of the LSP is automatically avoided in 
this model~\footnote{The other possibility pointed out in footnote 7, where
$\mathbb{Q},\, \mathbb{L}$ live in the bulk, and 
$U,\, D,\, E,\, H_{1,2}$ are localized on
the brane, would yield the charged slepton, $\tilde{\ell}_R$, as the
LSP, with a mass 
$m^2_{\tilde{\ell}_R}=9\,\alpha_1 m^2_{\tilde{q}}/20\,\alpha_3\simeq 0.35\,
\mu$, while the sneutrino $\tilde{\nu}_L$ is heavy, with a mass 
$M_{\tilde{\nu}_L}\simeq \omega/R$.}.

\section{Unification and non-perturbativity scales}

In this section we will comment on the sensitivity of the model to the sector
of the theory beyond the scale $1/R$. It is a well known fact~\cite{vt}
that, because of the non-renormalizability of the 5D theory, the gauge and
Yukawa couplings run with a power law behaviour of the scale. This idea was
on the basis of the so-called 
``accelerated'' unification proposed in Ref.~\cite{dienes}
and subsequently analyzed in different papers~\cite{graham}-\cite{manel}.
In particular, the gauge and Yukawa coupling 
one-loop renormalization for the
model analyzed in this paper were studied in Refs.~\cite{dq1,dq2} with
one-loop $\beta$-functions:
\begin{align}
16\pi^2\beta_{g_1}=&\left(\frac{48}{5}e^t-3\right)g_1^3 \nonumber\\
16\pi^2\beta_{g_2}=&\left(-4\,e^t-3\right)g_2^3 \nonumber\\
16\pi^2\beta_{g_3}=&-3\,g_3^3 \nonumber\\
16\pi^2\beta_{h_t}=&\left\{e^t\left(4 h_t^2+h_b^2-3 g_2^2-\frac{1}{3}
g_1^2-\frac{8}{3}g_3^2\right)+2 h_t^2-\frac{8}{15} g_1^2-\frac{8}{3}
g_3^2 \right\} h_t\nonumber\\
16\pi^2\beta_{h_b}=&\left\{e^t\left(4 h_b^2+h_t^2-3 g_2^2-\frac{1}{3}
g_1^2-\frac{8}{3}g_3^2\right)+2 h_b^2-\frac{2}{15} g_1^2-\frac{8}{3}
g_3^2 \right\} h_b
\label{rge}
\end{align}
where $g_2\equiv g$ and $g_1\equiv\sqrt{5/3}\,g'$ are the 
$SU(2)_L\times U(1)_Y$ gauge couplings with hypercharge normalization 
$k_1=5/3$.

We have run the one-loop
renormalization group equations (RGE) from $M_Z$ to scales
$\mu> 1/R$ using, for $M_Z\leq \mu\leq M_{\rm SUSY}$ the SM beta functions
for gauge and Yukawa couplings~\footnote{To simplify the analysis we 
use here a common scale of supersymmetry breaking 
$M_{\rm SUSY}\simeq 1$ TeV.}, those of the MSSM for 
$M_{\rm SUSY}\leq \mu\leq 1/R$, and those in Eq.~(\ref{rge}) for scales
$\mu> 1/R$. We have chosen for the plot 
$t_\beta\simeq 37$ and $1/R\simeq 4$ TeV.

\begin{figure}[htb]
\centering
\vspace{5mm}
\epsfig{file=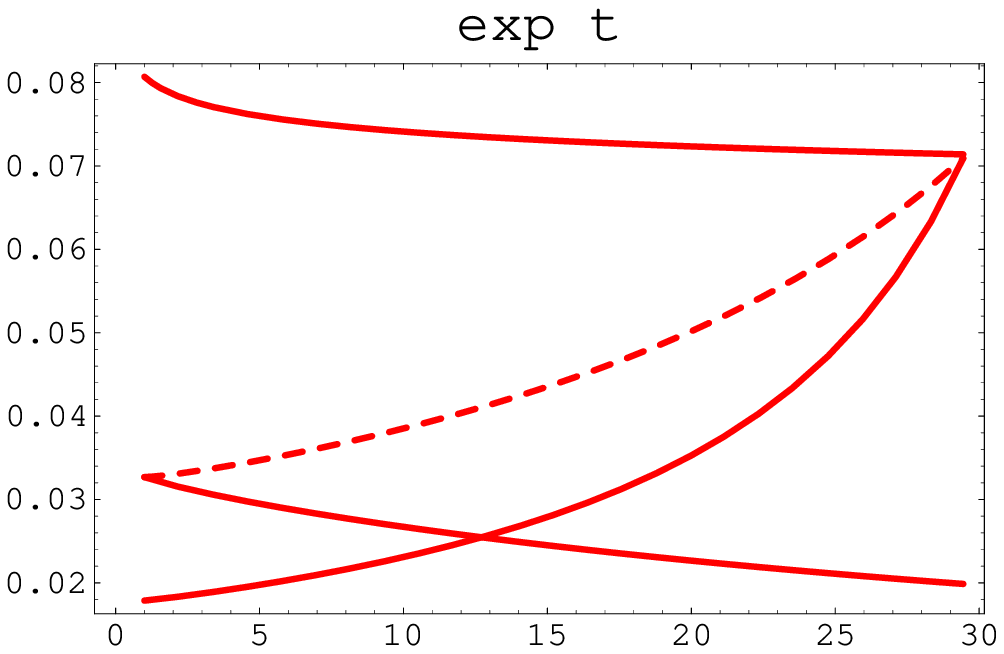,width=0.48\linewidth}\hspace{.3cm}
\epsfig{file=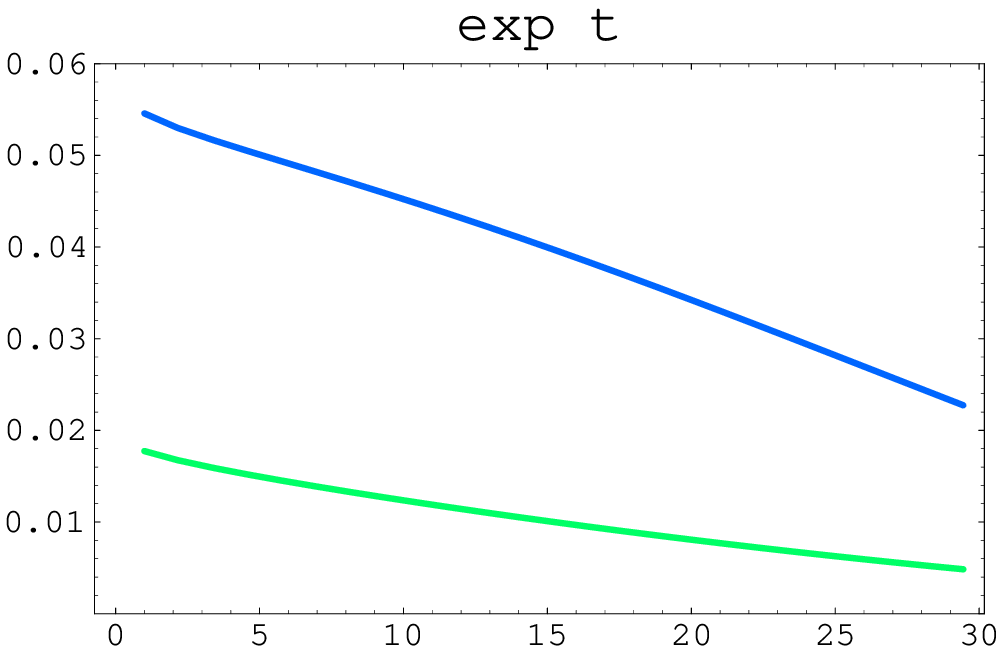,width=0.48\linewidth}\hspace{.3cm}
\caption{Plot of the gauge couplings $\alpha_i$, $i=1,2,3$ (left panel),
$i=t,b$ (right panel) as a function of $\exp(t)\equiv R\,\mu$.}     
\label{Fig:running}
\end{figure}

The result is shown in Fig.~\ref{Fig:running} where we plot
the gauge couplings $\alpha_i=g_i^2/4\pi$, $i=1,2,3$ (solid curves in the
left panel) and the Yukawa couplings $\alpha_{t,b}=h_{t,b}^2/4\pi$ 
(right panel). We see that all couplings, except $\alpha_1$, are 
asymptotically free and, as a consequence, in the region $\mu R\simlt 30$
shown in the plots the theory is weakly coupled. For
$\mu R> 30$, $\alpha_1$ keeps on growing and for $\mu R\simeq 40$ the theory
becomes strongly coupled.

We can compare here the above scales with those obtained in 
models~\cite{Barbieri,Hall} where
all SM particles are propagating in the bulk. There are two main differences:
\begin {itemize}
\item
The first important difference concerns the running of the QCD gauge coupling,
$g_3$. In our case there is no linear running in $\beta_{g_3}$ because only
half of the SM $SU(3)$ triplets (the right-handed ones) are propagating in
the bulk which leads to cancellation of the coefficient of the linear term
in $\beta_{g_3}$. When also the left-handed triplets propagate in the bulk
there is an extra contribution to the $SU(3)$ $\beta$-function as
$16\pi^2 \Delta \beta_{g_3}=6 e^t g_3^2$ which makes $g_3$ to increase with
the scale and become non-perturbative at $\mu_{\rm NP}\,R\simeq 6-8$, 
depending on the value of 
$1/R$~\footnote{Since $SU(3)$ is asymptotically free in the SM,
the smaller $1/R$ the larger $\alpha_3(1/R)$ and the smaller $\mu_{NP}\,R$.}. 
However in our case we have seen that only $\alpha_1$ is non-asymptotically
free and has a linear running, which translates into a much larger 
non-perturbative scale $\mu_{\rm NP}\,R \simeq 40$.

\item
The second important difference concerns the running of Yukawa couplings.
As a consequence of the superpotential structure $UTT$ 
the $\beta$-functions of the
Yukawa couplings are governed by the anomalous dimensions of fields in the
brane. These anomalous dimensions involve a single $U$-field propagating in the
loop, which makes their scale dependence linear. On the other hand,
theories with all matter fields propagating in the bulk rely on localized
Yukawa couplings with a superpotential structure of the type $UUU$. In that
case the one-loop anomalous dimensions involve two $U$-fields propagating in
the loop, which makes the scale dependence Yukawa $\beta$-functions 
quadratic and
worsens the perturbative behaviour of Yukawa couplings. In fact we have seen
that, while in our case the top and bottom Yukawa couplings are one-loop
asymptotically free, in theories with all SM fields in the bulk they become
non-perturbative at scales $\mu_{\rm NP}\,R \simeq 3-6$~\cite{Barbieri,Hall}.
\end{itemize}

Finally we can see from the left plot of Fig.~\ref{Fig:running} that the 
theory does not unify. This issue was analyzed in Ref.~\cite{dq1} where it
was shown that by adding two zero hypercharge triplets~\footnote{In the spirit
of Refs.~\cite{triplets}.} propagating in the bulk, and contributing to
the $\beta$-functions 
$16\pi^2\Delta \beta_{g_2}=8 e^t g_2^3$, the theory unifies
at $\mu_{\rm GUT}\,R\simeq 30$. The
running of $\alpha_2$ is shown in Fig.~\ref{Fig:running} (left pannel) in
dashed where unification is explicit. 
Of course sensitivity with the scale is large and to draw firm
conclusions on unification predictions we should control all threshold
effect at the scale $\mu_{\rm GUT}$ from the underlying (string) theory.

\section{Conclusions}

In this paper we have presented a 5D model with the fifth dimension 
compactified on the orbifold $S^1/\mathbb{Z}_2$, where the parity 
$\mathbb{Z}_2$ is defined by $x_5\to - x_5$ and an appropriate lifting to
spinor and $SU(2)_R$ indices. The gauge bosons are propagating in the bulk
($U$-states) while matter and Higgs fields can either propagate in the bulk or
be localized on the 3-branes at the orbifold fixed points ($T$-states). The
orbifold selection rules allow superpotential interactions 
on the branes of the type
$UUU$ and $UTT$, where the former are suppressed as $(\Lambda_s R)^{-1}$ with 
respect to the latter. We then assume a superpotential of the form $UTT$.

Radiative finite electroweak symmetry breaking is triggered by right-handed
stops propagating in the bulk, where supersymmetry is broken by a
Scherk-Schwarz mechanism that uses the $U(1)_R$ subgroup of the
$SU(2)_R$ ($R$-symmetry) left unbroken by the compactification.
The Scherk-Schwarz parameter $\omega$ is considered as a 
free parameter in this paper.
To allow fermion masses and superpotential
interactions on the brane, consistent with the orbifold selection rules, we
must assume that $SU(2)_L$ singlets propagate in the bulk while $SU(2)_L$
doublets are localized on the 3-brane, e.g. at the fixed point $x_5=0$.
In this way the 4D theory of zero modes and localized states constitute 
the MSSM ($N=1$), while that of massive modes possesses $N=2$ 
supersymmetry in four dimensions.

The Higgs sector is localized on the brane and then
the $\mu$-parameter does not arise
through compactification, although it is an allowed term in
the superpotential. One has to consider it as an effective parameter from the
underlying, supergravity or string, theory. In this sense the situation is
better than in the MSSM since the cutoff $\Lambda_s$ is in the TeV range.

We have proven that electroweak breaking is induced by finite radiative
corrections triggered by the top/stop sector, $\mu$ and $\omega$ 
remaining as the only free parameters. In fact, the compactification radius and
$\tan\beta$ are fixed by the minimization procedure. While $\tan\beta$ is
large, $\tan\beta\sim 40$, $1/R$ is in the range, $1-15$ TeV, depending on
the values of $\mu$ and $\omega$. The lightest Higgs mass is similar
to the MSSM one~\cite{massold} for large $t_\beta$, in particular for 
large values of $1/R$ for which the heavy KK-modes can be integrated out. 
We have checked that their integration produces threshold effects that 
increase the lightest Higgs mass by an amount $\simlt 5$ GeV so we expect
that once we include the genuine MSSM two-loop 
corrections~\cite{massnew} the final value
predicted for the model will not differ much from the MSSM one.
On the other hand LEP searches on the MSSM Higgs sector imply 
($\omega$ dependent) lower bounds on the $\mu$ parameter and the
mass of the SM-like Higgs, $M_h$. They translate into the absolute
bounds $\mu\simgt 350$ GeV and $M_h\simgt 145$ GeV.

The supersymmetric mass spectrum has a well defined pattern. 
The heaviest states are right-handed sfermions, the gauginos and the 
gravitino, with a mass $\sim\omega/R$, and higgsinos, with a mass $\sim\mu$. 
The other supersymmetric partners (left-handed sfermions) receive radiative 
masses from the supersymmetry breaking in the bulk. The next to 
heaviest states 
are the left-handed squarks which receive radiative masses from the
gluon/gluino sector. We have found that the lightest supersymmetric particles
are the sneutrinos, and the next to lightest supersymmetric particles 
the charged sleptons, with a squared mass difference between them $\sim M_Z^2$.

The gauge and Yukawa couplings run linearly with the scale $\mu$
(a reflection of the non-renormalizability of the 5D theory). 
The theory remains perturbative for scales $\mu\, R\simlt 40$ while
$\alpha_1$ and $\alpha_3$ unify at $\mu\, R\simeq 30$. This is a great
difference with respect to recently proposed models, where all particles
propagate in the bulk~\cite{Barbieri,Hall} and that rely on superpotential
interactions of the type $UUU$. In this case wave function renormalization
involves one-loop diagrams with two bulk states propagating and vertices that
do not conserve the KK-number. This translates into quadratic running for
the Yukawa couplings and make the theory non-perturbative for 
$\mu\, R\simeq 3-6$. Since all properties of the theory rely on summing 
over the infinite tower of KK states, having such a low cutoff is a 
drawback which can make threshold effects from heavy KK-modes
non-negligible.

Concerning the recent related works already mentioned, 
Ref.~\cite{Barbieri} uses a single Higgs
hypermultiplet in the bulk and all matter fields propagating in the bulk.
It breaks $N=1$ supersymmetry by a SS-mechanism 
using $R$-parity as the global symmetry,
which amounts to choosing $\omega=1/2$ as the SS-parameter. On the one hand,
since they have a single Higgs field, they solve the $\mu$-problem by 
nullification. On the other hand, they have to rely on $UUU$ Yukawa couplings
to give masses to quarks and leptons, which makes, as we noticed above, the
theory non-perturbative for low scales. The prediction on the Higgs mass
($\sim$ 128 GeV) is reached by our model for the particular case $\omega=1/2$,
$\mu=0$, although this is a non-physical point for our theory. Finally the
LSP in Ref.~\cite{Barbieri} is the stop, also making a big difference with
respect to our theory which predicts the sneutrino as the LSP.

Concerning Ref.~\cite{Hall}, its authors 
use two Higgs multiplets (either in the bulk
or localized on the brane) and also all matter fields propagating in the bulk.
Supersymmetry is broken by the SS-mechanism using $R$-parity, i.e. again
$\omega=1/2$~\footnote{Actually Ref.~\cite{Hall} also proposes another, 
non-SS mechanism for supersymmetry breaking based on a 
strong supersymmetry breaking localized on a hidden brane that has 
little relation with that used in this paper and that we will not comment on
here.}. Since massive gauginos form Dirac fermions, they cannot generate
radiatively the $m_3^2\,H_1\cdot H_2$ term in the potential and 
this model has to rely on non-renormalizable terms in the superpotential, 
as $\lambda\left(H_1\cdot H_2\right)^2/\Lambda_s$ that, 
along with the $\mu$-term,
generate the $m_3^2$-parameter, which then gets a sensitivity to the
cutoff scale, at the tree level. Also Yukawa couplings
renormalize quadratically with the scale and so the theory becomes 
non-perturbative at low scales.

Let us finally conclude by emphasizing the beauty of the mechanism of 
electroweak breaking triggered by a top-quark propagating in the bulk of 
an extra dimension. In our opinion it constitutes a step forward in our 
understanding of electroweak breaking since the generated Higgs field 
instability at the origin is neither ``imposed'' by hand at tree-level 
nor sensitive to the Standard Model cutoff, in spite of its radiative origin. 
Of course the top-quark propagation in the bulk is not without
phenomenological consequences~\cite{paco}, 
that should be worked out in future works and
must be tested in future colliders. On the other hand, 
the pattern of supersymmetric mass spectrum should give a hint on the 
particular mechanism of supersymmetry breaking realized by the Nature.

\section*{Acknowledgements} 
We thank Jos\'e Ram\'on Espinosa and Carlos Wagner for discussions
concerning present LEP bounds and
Alex Pomarol for useful discussions and for participating in the 
early stages of this paper. The work of AD was supported 
by the Spanish Education Office (MEC) under an \emph{FPI} scholarship.   

\end{document}